\documentclass[
reprint,
amsmath,amssymb,
]{revtex4-1}

\usepackage{graphicx}
\usepackage{dcolumn}
\usepackage{bm}
\usepackage{float}
\usepackage[font={small,it}]{caption}

\begin{document}

\title{Switching between normal and anomalous Laser Induced Periodic Surface Structures\\
}

\author{Ihor Pavlov$^1$$^,$$^2$$^,$$^*$, Ozgun Yavuz$^3$, Ghaith Makey$^1$, Onur Tokel$^1$ and Omer Ilday$^1$$^,$$^3$$^,$$^4$}

\affiliation{$^1$Department of Physics, Bilkent University, 06800 Ankara, Turkey
\\
 $^2$Institute of Physics of the NAS of Ukraine, Nauki av. 46, 03028 Kyiv, Ukraine
 \\
$^3$Department of Electrical and Electronics Engineering, Bilkent University, Ankara, 06800, Turkey
\\
$^4$UNAM – National Nanotechnology Research Center and Institute of Materials Science and Nanotechnology, Bilkent University, Ankara, 06800, Turkey
\\
$^*$pavlov.iop@gmail.com
}

\begin{abstract}
We report on the studies of switching mechanism between normal and anomalous laser induced periodic surface structures. We have shown that for high loss metals the switching mechanism between normal and anomalous modes relays on an interplay between two different feedbacks inherent into the structure formation process: long range, low intensity dipole-like scattering of light along the surface, which governs anomalous ripples parallel to the laser polarization, and short range, high intensity plasmon-polariton wave, which is initiated by near field dipole radiation and responsible for creation of ripples perpendicular to the polarization, i. e. normal structure. By managing these two feedbacks, we demonstrated creation of both normal and anomalous laser induced periodic surface structures on the same surface. In contrast to the previous studies, we have shown that the thermal oxidation mechanism can form both normal and anomalous types of the structure, while the ablation mechanism is involved only during the normal structure formation. Unlike the formation of oxidation type anomalous structure, the formation of ablation type normal structure does not have inherent negative feedback, which is self-regulating nano-ripples formation. This feedback needs to be applied artificially, by regulating scanning speed and pulse energy at a given repetition rate of the laser. With implementation of nonlinear laser lithography technique for ablation type of normal laser induced periodic surface structures we demonstrated large area high regular nano-grating formation on different surfaces with extremely high industrially acceptable speed.
\end{abstract}

\maketitle

Nowadays there is a significant growth in studies of Laser Induced Periodic Surface Structuring (LIPSS) \cite{1} due to both practical and theoretical importance, especially after establishing of Nonlinear Laser Lithography (NLL) technique \cite{2}, which opens new area for industrial application of LIPSS as an effective tool for controllable, highly ordered large area nanostructuring. Usually the structure appear on the surface under laser beam in the form of periodical lines (ripples) with close to the laser wavelength or sub- wavelength period, perpendicular or parallel \cite{3} to the laser polarization. (Note, in this article we are not studying so called high spatial frequencies LIPSS, which period is much smaller the wavelength of laser light.) The structure, which appear perpendicular to the laser polarization is called $"$normal$"$ in most of the literature in opposite to the structure parallel to the laser polarization, which is called $"$anomalous$"$ \cite{4}. Generally accepted basic mechanism for the normal structure creation is an interference between incident laser beam and plasmon-polariton wave induced and propagating along the surface \cite{5, 6, 7}. The basic mechanism for the anomalous structure creation is less generally accepted. In 70-th it was referred to an interference between the incident beam with $"$surface scattered waves$"$ \cite{8}, where every natural defect on the surface is scattering the light. However, since TE mode cannot propagate along the interface between two media as it does not satisfy boundary conditions of Maxwell`s equations, later this type of the structure was referred to an interference of incident light with so called $"$radiation remnant$"$ field \cite{9,10}. In \cite{2}, the anomalous structure was investigated by numerical simulation based on an interference of incident laser beam with the light scattered along the surface, were every small natural defect of the surface produce dipole-like scattered field. In addition to the good agreement with all previous experimental results, the simulation predicts a possibility of coherent extension of the structure over large area, by scanning the laser beam with a small diameter over the surface, which was proven experimentally. In this case, the small beam diameter preserves coherency of the scattered light within the beam spot, and the scanning introduces positive nonlocal feedback from already created structure to the new area on the surface. It is important to notice, although many papers demonstrated creation of the structure on the same material (for example on Ti) where it is normal (see for example \cite{3}) or anomalous \cite{2,11}, there is no any clear theoretical model presented in literature, which can explain the switching mechanism between these two modes. It was noticed already, that the formation of anomalous structure is associated with thermochemical oxidization, while the formation of normal structure is related to ablation \cite{12}. However, the internal physical mechanism was not discussed. 
In present paper we have shown that the switching mechanism between normal and anomalous modes relays on interplay between two different feedbacks inherent into the structure formation process: long range, low intensity dipole-like scattering of light near the surface, which governs anomalous ripples parallel to laser polarization, and short range, high intensity plasmon-polariton wave, which is initiated by near field dipole radiation and responsible for creation the structure perpendicular to polarization, i. e. normal ripples. In addition, we have shown, that both normal and anomalous structures can be formed by thermochemical oxidization, while the ablation is involved only in certain case of normal structure. 
\par

Firstly, we will investigate the creation of the structure with numerical simulation. The simulation of the structure formation process is based on two steps \cite{2}. The first step is calculation the light intensity distribution on the surface as result of an interference of the incident laser light with the light scattered from all defects of the surface. The second step is recalculation the surface morphology according to the intensity distribution. The inhomogeneous intensity distribution on the surface creates inhomogeneous temperature distribution, which triggers local oxidization or ablation in different points of the surface depending on the local temperature. These surface modifications will act as the new scatters during the next iteration.
In order to calculate the scattered light distribution along the surface we can assume that every defect of the surface (natural or already created by laser) scatters dipole like, as long as the size of the defect is match smaller than wavelength of the light \cite{13}.

Assuming, the incident light polarization is directed along the $x$ axis, the general equation for $E_x$ component of the scattered light including far and near field dipole zones can be written as \cite{14}:
\begin{eqnarray}
E(x,y)_x^x={\gamma}E_{0x}(x^\prime, y^\prime)
{h(x^\prime, y^\prime)} e^{ikr} \times\nonumber\\
\times
\Big[
\frac{cos^2{\theta}}{r}
+(3sin^2{\theta}-1)\Big(
\frac{1}{k^2r^3}
-\frac{i}{kr^3}
\Big)
\Big],
\end{eqnarray}
where $(x, y)$  is the position of the observation point and $(x^\prime, y^\prime)$ is the position of the scattering point; $\gamma$ is  polarizability ($\alpha$) dependent coefficient: ${\gamma}=\frac{{\alpha}k^2}{4{\pi}{\epsilon}_0}$; $E_{0x}(x^\prime, y^\prime)$ - is the $x$ component of the electrical field of incident light at the position $(x^\prime, y^\prime)$;  $\theta$ - is the angle between the polarization direction and the radius vector from the observation point to the scattering defect ($\theta=atan(\frac{y-y^\prime}{x-x^\prime})$); and $r$ - is the absolute value of the radius vector between observation point and scattering point. ${h(x^\prime, y^\prime)}$ is the local height of the surface (defect) at the $(x^\prime, y^\prime)$ position. The term $e^{ikr}$ represents a phase shift during the time propagation of scattered light from point $(x^\prime, y^\prime)$ to the point $(x, y)$, where $k=2\pi/\Lambda$ is the wavenumber of scattered light. 

The first term inside of rectangular brackets in the equation (1) corresponds to far zone field distribution. For the far field component, the direction of electrical vector $\bf E$ is parallel to the surface and perpendicular to propagation direction. It is well known, that such mode cannot propagate as a wave along the air/metal interface, since it does not satisfy Maxwell`s equations. However it can be taken into account  as a time-varying far zone non-radiative field, thus modulating the intensity of the incident light on the surface \cite{15}. The second term inside of rectangular brackets of the equation (1) corresponds to near zone field distribution. It is known, that the near field excites surface plasmon-polariton (SPP) wave, propagating in both directions from the defect along E vector of the incident light. Although, the quantitative description of SPP excitation is quite complicated, since it requires an integration of both $z$ and $y$ components of the field over all excitation angles \cite{16}, for the aim of our studies we can substitute the integration by introducing coupling constant $\xi$ between near field components and SPP. Since the SPP decays exponentially with a distance from the source, the resulting equation can be written as follows:
\begin{eqnarray}
E(x,y)_x^x={\gamma}E_{0x}(x^\prime, y^\prime)
{h(x^\prime, y^\prime)} e^{ikr} \times\nonumber\\
\times
\Big(
\frac{cos^2{\theta}}{r}
+{\xi}sin^2{\theta}e^{-{\beta}r}
+B
\Big),
\end{eqnarray}
where $\beta$ is SPP decay constant, and $B=\frac{1+ikr}{k^2r^3}$, which is polarization independent term.

Similarly, we can write the equations for $E(x,y)_y^x$ component of scattered field for the incident light polarized along $x$ direction, and $E(x,y)_x^y$, $E(x,y)_y^y$ components of scattered field for the incident light polarized along y direction. In order to calculate the result intensity distribution for each point of the surface, we integrate every component of the scattered field from all points of the surface:

\begin{eqnarray}
E(x,y)_x^x={\iint}\Big[{\gamma}E_{0x}(x^\prime, y^\prime)
{h(x^\prime, y^\prime)} e^{ikr} \times\nonumber\\
\times
\Big(
\frac{cos^2{\theta}}{r}
+{\xi}sin^2{\theta}e^{-{\beta}r}
+B
\Big)\Big]{dx^\prime}{dy^\prime},
\end{eqnarray}

Finally, for every point of the surface we are summing the components of the incident light with the corresponding components of the scattered light, and calculating the local intensity for every point of the surface which defines the local pulse fluence at a given pulse duration. It allows as to calculate the local temperature at every point of the surface, assuming that the pulse duration is much smaller than the heat dissipation time.
\\
Fig. 1a demonstrates the simulated intensity distribution on Ti surface around a single TiO$_2$ spherical defect with 100 nm diameter.
Depending on the metal properties, the range of plasmon-polariton wave varies from sub-micron for high loss metals with low (negative) real part of the relative permittivity of the metal, and a large imaginary part (Ti, Cr, W etc.), to several tens microns for low loss metals (Au, Ag, Cu) with large (negative) real part and a low imaginary part of the relative permittivity \cite{17}. For Ti, which we are using in our simulation, the SPP propagation range is 0.2 $\mu$m. In opposite, for the far dipole zone, the electrical field on the metal surface is zero for ideal conductor, and non-zero for high loss metals, which can be estimated from \cite{18}. Fig. 1b demonstrates the cross section of intensity distribution along $x$ and $y$ direction from the Fig. 1a. Along the $x$ direction, which is the direction of SPP propagation, there are two strong peaks near the defect with a fast decay due to short SPP length. Along the $y$ direction, although the peak intensity is lower, there is a long range intensity modulation due to longer range of the far zone field. Such intensity distribution causes inhomogeneous heating of the surface, or even ablation for the places, where the local pulse fluence reaches ablation threshold. If the Ti sample is placed in air (which is true for the most of experiments performed up to now), the temperature dependent oxidation will start right after the laser pulse, which modifies original height of every point of the surface. The dependence of the changes of the local height $\Delta$h (related to oxidation rate) per pulse, on local pulse fluence is shown on Fig. 1c. For the local pulse fluence below 300 $J/m^2$, which corresponds to the local surface temperature below 1000 degree the data are recalculated from \cite{19}. For the range between 370 $J/m^2$ to 1170 $J/m^2$ the oxidation rate is extracted from experimental data presented on Fig. 1d (see Appendix A for more details). For these two ranges the local growth of Ti oxide is exponentially increases with the local pulse fluence, and the $\Delta$h is in order of $10^-$$^6$   - $10^-$$^3$ nm/pulse. For the range above 1470 J/m$^2$ (above the calculated ablation threshold), the local height decreases due to the ablation, which we calculated in the assumption, that all energy absorbed in the skin depth is spent to heat, melt, and evaporate the volume. In this range, the local changes of $\Delta$h is much higher, and corresponds to 40-50 nm/pulse, depending on the local pulse fluence. It is obvious, that the created structure in this case is $"$negative$"$ image of the oxidation type, since it consists of deepening instead of bumps on the surface.

\begin{figure}[b]
\includegraphics[scale=0.4]{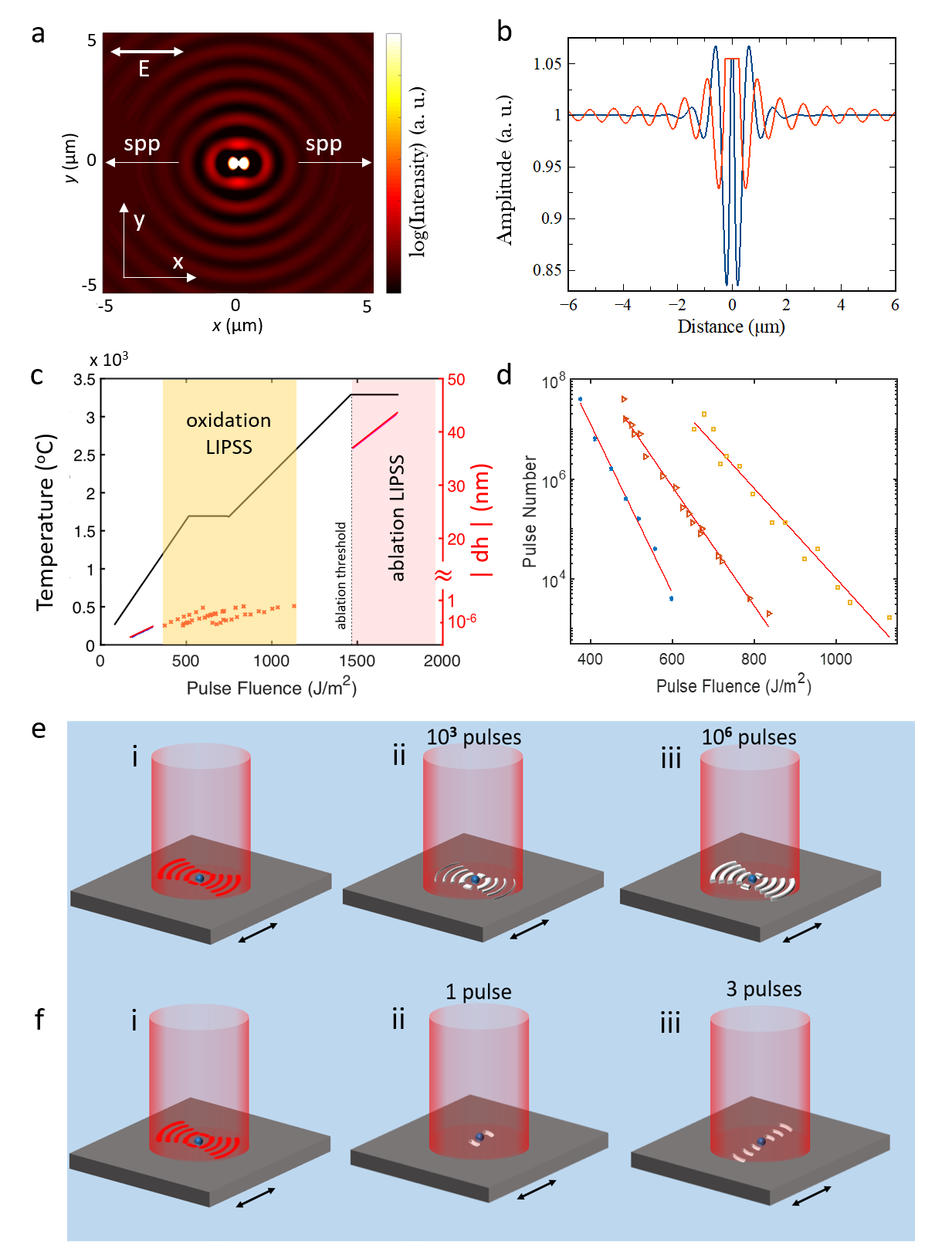}
\caption{\label{fig:epsart} 
(a) Intensity distribution as a result of interference of incident light with light scattered from single TiO$_2$ defect placed on Ti surface. Two arrows indicate surface plasmon-polariton propagation direction. (b) $x$ (black line) and $y$ (red line) cross sections of the intensity distribution from (a). (c) The local temperature of skin depth of Ti as a function of local pulse fluence (black solid line, left axis), absolute value of local height changes as a function of single pulse fluence (right axis) recalculated from the literature (blue line), obtained from experiment (red crosses) (see explanation in text) and calculated for the pulse fluence above the ablation threshold (magenta solid line).  (d) The number of pulses as a function of pulse fluence required to produce anomalous LIPSS at different repetition rate: 2 MHz (dots), 1 MHz (triangles), 0.5 MHz (squares), the red lines are exponential fit to the data. (e, f) schematic cartoon demonstrating formation of: (e) anomalous oxidation LIPSS after $10^3$ pulses (ii), and $10^6$ pulses (iii), (f) normal ablation LIPSS after 1 pulse (ii), and 3 pulses (iii). The initial field distribution (i) is the same in both cases.}
\end{figure}

Qualitatively, the formation process of normal and anomalous structures is shown on Fig. 1e, 1f. If the sample is illuminated with the pulse fluence significantly below the ablation threshold (Fig. 1e), after each pulse small oxidation occurs around every defect mostly in the places with maximum intensity. As the intensity modulation has longer range in $x$ direction, more oxidation lines are created parallel to the polarization. Since every new oxidation point, which will act as a new scatter for the next pulse, is located on the distance equal the wavelength of scattered light from each other, this positive feedback will establish coherent growth of anomalous structure. This process is slow, it requires thousands of pulses, and normally observable for high repetition rate fs laser. However, if the sample is illuminated with the pulse  fluence near the ablation threshold (Fig. 1f), the places with highest intensity (first peaks of SPP on Fig. 1b) will exceed the ablation threshold. In this case, every new pulse creates new lines oriented perpendicular to the laser polarization, governing normal LIPSS.

The numerical simulation result demonstrating creation of normal and anomalous structures is shown on Fig. 2a (i) and 2a (ii) correspondingly. For the Fig. 2a (i) we raster scanned the surface with the pulse fluence below the ablation threshold with several thousand pulses per spot. For the Fig. 2a (ii) we raster scanned the surface with the pulse fluence near the ablation threshold and the speed corresponding to several pulses per spot. The results clearly demonstrate switching from anomalous to normal types of LIPSS with increasing the pulse fluence above the ablation threshold. 

In our experiments we used home-built Yb femtosecond laser operating at 1030 nm \cite{20}. At 1 MHz repetition rate, the maximum pulse energy and the shortest pulse duration are 1 $\mu$J and 100 fs correspondingly. The repetition rate of the laser was electronically switchable from 40 MHz (the repetition rate of the oscillator) to single pulse operation. The laser beam was coupled into high resolution optical microscope in both transmission and reflection mode. In transmission mode we can directly observe the structure formation in real time on thin metal films deposited on glass substrate. In reflection mode we can observe the structure formation in real time on thick non transparent metal films or on the bulk samples.


Additionally, one branch of the laser beam was coupled to high speed galvo-scanner, which allows fast creation of the structure on large area samples.

\begin{figure}[b]
\includegraphics[scale=0.65]{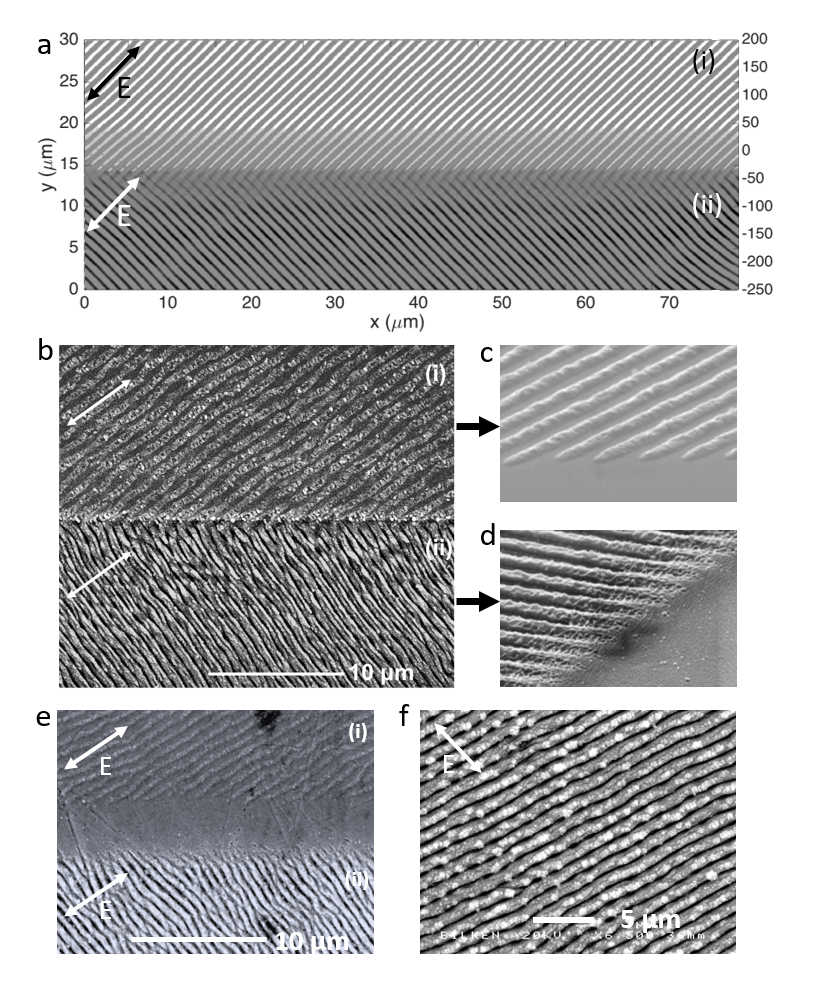}
\caption{\label{fig:epsart} 
(a) numerical simulation result for 1000 pulses per spot with pulse fluence below the ablation threshold (i) and several pulses per spot with the pulse fluence above the ablation threshold (ii). (b-e) SEM images of normal and anomalous LIPSS created on Ti surface. (b) anomalous oxidation LIPSS created with low pulse energy, high repetition rate laser (i), and normal ablation LIPSS created with high pulse energy ''low'' repetition rate laser (ii) (see explanation in the text). (c) tilted SEM view of anomalous oxidation LIPSS (from b(i)). (d) tilted SEM view of normal ablation LIPSS (from b(ii)). (e) anomalous LIPSS created on Ti surface with low pulse energy, high repetition rate laser (i) (the same as b(i)), and normal ablation LIPSS created on Ti surface with low pulse energy, high repetition rate laser by suppressing oxidation (ii). (f) normal oxidation LIPSS created on copper surface with low pulse energy, high repetition rate laser.
}
\end{figure}

Fig. 2 (b-e) demonstrate SEM images of different types of the structure created on thick (400 nm) Ti film deposited on Si substrate. On the top part of Fig. 2b (i) the sample was raster scanned with the pulse fluence 700 $J/m^2$, which is below the ablation threshold. The scanning speed was 50 $\mu$m/s, which corresponds to nearly $6{\cdot}10^4$ pulses per spot (the spot size was 3 $\mu$m). In this case we observe anomalous structure creation with the groves direction parallel to the polarization of light. In the bottom part of Fig. 2 b(ii) the laser is switched to single pulse mode and after every step we shift the sample to 1 $\mu$m, which corresponds to ~3 pulses per spot. In this mode we again raster scanned the beam over the sample. The single pulse fluence in this case was ~1600 $J/m^2$, which is above the ablation threshold. In this case we observe creation of normal structure, with the groves direction perpendicular to the polarization of light. The SEM tilted view demonstrates growth of the anomalous structure above the surface level (Fig. 2c), which corresponds to oxidation, while the normal structure consists of periodical deepening, which corresponds to ablation (Fig. 2d).
 
Thus, different types of LIPSS can be created on Ti surface by manipulating two different positive feedback mechanisms, one of which is governed by oxidation in the direction of long range far zone dipole field, and another one is governed by ablation from near zone dipole field coupled into SPP. One way of such manipulation, which we already demonstrated, is increasing pulse fluence above ablation threshold. Another possibility of positive feedback control is suppressing oxidation rate, which governs the anomalous structure creation. Such demonstration is shown on the Fig. 2e. On the top part of the image (Fig. 2e (i)) we created anomalous LIPSS pattern with laser parameter and scanning conditions similar to described for Fig. 2 b(i). On the bottom part of the image (Fig. 2e (ii)), the laser parameters and scanning conditions are the same, however we applied Ar gas flow to processed zone in order to suppress oxidation rate. At such conditions the positive feedback from the oxidation is suppressed and the direction of groves is switched from parallel to perpendicular to the laser polarization.

Up to now we demonstrated on Ti that the anomalous structure, i. e. the structure which groves are parallel to the polarization of the incident light, consists of oxidation lines, while the normal structure consists of elongated ablation craters. However, for low loss metals, where the SPP range is long enough to dominate over the dipole far zone field, only one direction of the structure is possible for both oxidation and ablation based grooves. This direction is perpendicular to the polarization of incident light. As an example, we demonstrated the oxidation based structure, where the groves are perpendicular to the laser polarization (normal) on copper film (100 nm) deposited on glass substrate (Fig. 2f).  In this case, the pulse fluence was 3500 $J/m^2$, which is much below the reported ablation threshold for copper 7900 $J/m^2$ \cite{21}. The effective number of pulses per spot was $10^5$. The SPP propagation length for copper is calculated to be ~65 $\mu$m.

As demonstrated above, there are two types of LIPSS is possible to obtain on the surface of high loss metals, by manipulating the positive feedbacks. With low pulse fluence (below the ablation threshold), high repetition rate laser in air atmosphere we obtain anomalous oxidation LIPSS parallel to the polarization of the laser (Fig. 2b (i), Fig. 2e (i)). By suppressing the oxidation rate (Fig. 2e (ii)), or increasing the pulse energy to the ablation threshold (Fig. 2b (ii)) we can switch the structure formation to normal ablation LIPSS with the direction perpendicular to the polarization of the laser. For low loss metals, with long SPP propagation length both oxidation and ablation LIPSS are normal, i. e. perpendicular to the polarization of the laser, however, if there is no other limitations (for example, the metal should be able to oxidize) which make the structure creation impossible.

Recently introduced NLL method \cite{2} demonstrated extremely regular large area nano-pattern formation with anomalous oxidation LIPSS on Ti, Cr, and W. The NLL technique is based on coherent extension of the pattern by scanning small laser beam over the surface with partial overlapping with already existing structure. In this case, the laser beam located in every position of the surface forms interference pattern with the part of the beam scattered on previously formed structure. It introduces non-local positive feedback, which leads to self-regulating and self-healing process, thus improving the quality of the structure. Here we demonstrate the application of NLL to ablation type normal LIPSS. As was shown before for anomalous oxidation LIPSS \cite{2}, in addition to non-local positive feedback, which is responsible for self-regulating long range periodicity, there is a negative feedback, which cause saturation of the structure growth after it reaches a certain height. This feedback is based on self-limitation of oxygen access to the growing ripple with increasing the height of the ripple. Practically, it makes the NLL process insensitive to multiple scans, as well as removes the lower limit for the scanning speed. The SEM images of oxidation anomalous LIPSS formed by NLL technique on Ti surface are presented on the Fig. 3 (a-c), where the structure was created by raster scanning the beam  once (a), twice (b), and 5 times (c) over the same surface. Visually, there is no difference between these structures in terms of their quality due to negative self-saturating feedback. In contrast, the ablation normal LIPSS does not have inherent negative feedback to saturate the structure formation. Every next pulse ablates more and more material, finally ruining the structure if the number of pulses per spot significantly exceeds the optimal number. It makes the NLL process for ablation normal LIPSS sensitive to multiple scans, as well as it defines the lowest scanning speed for a given pulse energy and repetition rate of the laser. The SEM images of ablation normal LIPSS created by NLL technique on Ti surface with 1, 2, and 5 scans are presented of Fig. 3 (d), (e), and (f) correspondingly. Clearly, even after the second scan, the quality of the structure decreases, and it ruined dramatically after the five scans.
\begin{figure}[b]
\includegraphics[scale=0.25]{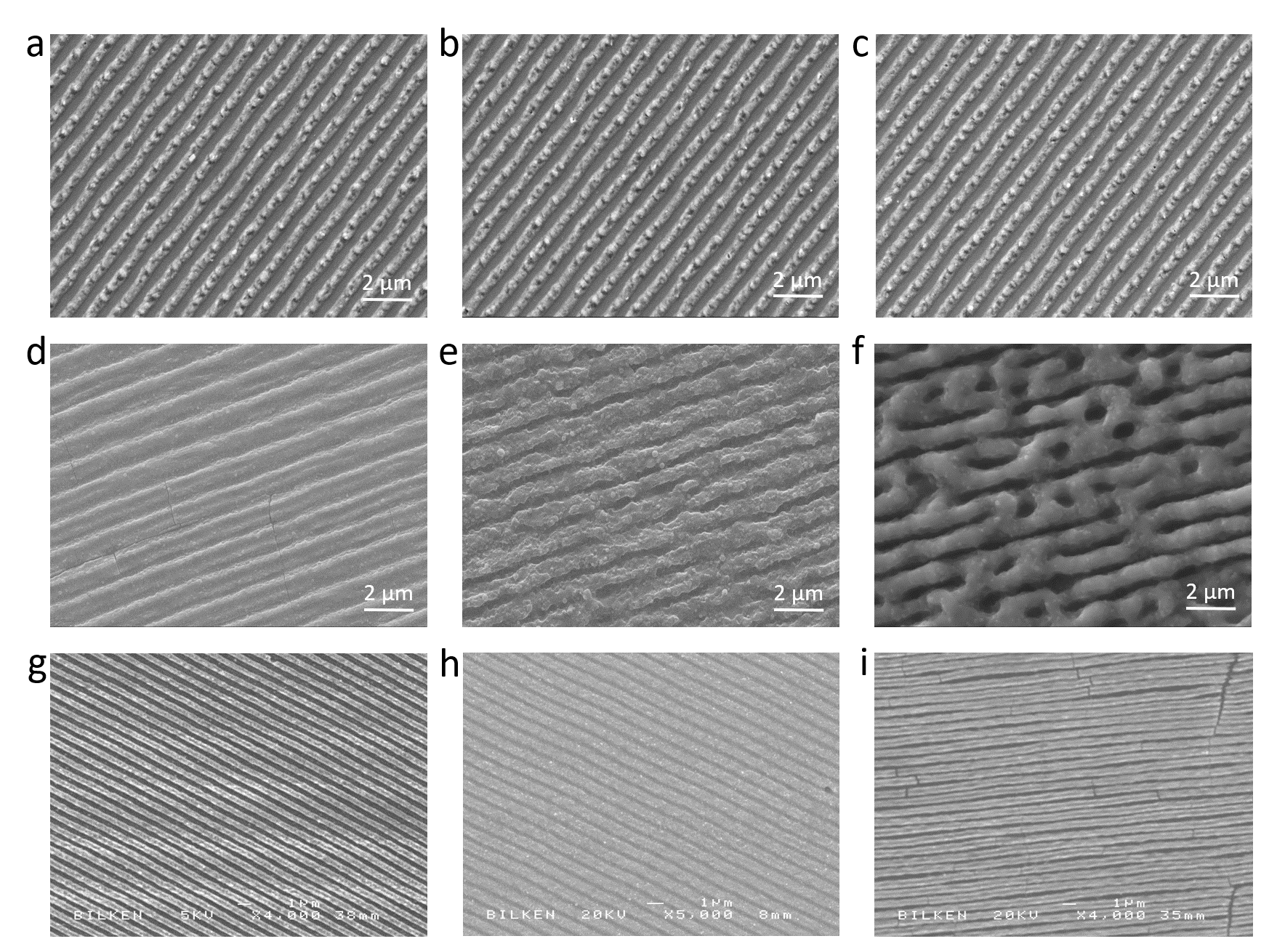}
\caption{\label{fig:epsart} 
SEM images of oxidation anomalous LIPSS (a-c) and ablation normal LIPSS (d-f) created by NLL technique on Ti surface with 1 scan (a, d), 2 scans (b, e) and 5 scans (c, f). (g-i) High regular LIPSS created by NLL on Ti (g), Cr (h), and Mo (i) surfaces.
}
\end{figure}
The absence of the natural negative feedback for ablation normal LIPSS, requires additional precautions to regulate the quality of the structure, while processing the sample with NLL technique. Since every pulse is creating the groves with a depth of 30-50 nm (Fig. 1 c), the optimal number of pulses per spot is 4-10, depending on the material properties. This requirement can be satisfied by regulating the scanning speed of the beam. For example, the scanning speed for the structure on Fig. 3 d, was 1500 mm/s, with the spot size ~11 $\mu$m at 1 MHz repetition rate of the laser, and 350 mW of average power. It corresponds to ~6 pulses per spot with 3 $kJ/m^2$ pulse fluence. With similar parameters, the NLL technique allows to create high regular nanostructure on many other materials. Fig. 3 (g-i) demonstrates highly regular nanostructures on Ti, Cr, and Mo. All of the samples were 300 nm thick films deposited on glass substrates. By taking into account high processing speed for ablation type normal LIPSS it opens a door for wide range of industrial applications. Even in our experiments, where we were limited by average power of the laser (which does not allow increase repetition rate of the laser by keeping the same pulse energy) and scanning speed of galvo-scanner (3000 mm/s), the processing speed for most of the surfaces was around 15 s for 1x1 $cm^2$ area. These parameters can be easily boosted for industrial laser processing setups.

As a conclusion, we have shown, that the switching mechanism between normal and anomalous LIPSS relies on interplay between two different feedbacks inherent into the structure formation process: long range, low intensity far zone dipole field, which governs anomalous ripples parallel to the laser polarization, and short range, high intensity plasmon-polariton wave, which is initiated by near zone dipole field and responsible for creation of the structure perpendicular to the laser polarization, i. e. normal ripples. By regulating these two feedbacks, at the first time to the best of our knowledge we demonstrated creation of both normal and anomalous structures on the same surface. In contrast to the formation of oxidation type anomalous LIPSS, the formation of ablation type normal LIPSS does not have inherent negative feedback, which is self-saturating nano-ripples formation. This feedback needs to be applied artificially, by regulating scanning speed and pulse energy at a given repetition rate of the laser. With implementation of NLL technique for ablation type normal LIPSS we demonstrated large area high regular nano-grating formation on different surfaces with extremely high industrially acceptable speed. We expect the obtained results will have impact on practical implementation of the method for different technological application.

\begin{acknowledgments}
We wish to acknowledge the funding from European Research Council (ERC) Consolidator Grant ERC-617521 NLL. We thank Dr. Coskun Kocabas, Dr. Alpan Bek and Dr. Tahir Colakoglu for fruitful discussions. We thank Dr. Tahir Colakoglu for SEM characterization performed in Center for Solar Energy Research and Applications of Middle East Technical University. We thank Murat Gure and Ergun Karaman for technical support of the work.
\end{acknowledgments}

\appendix
\section{}
In order to define experimentally the amount of Ti oxide created by a single pulse for anomalous oxidation structure, for a given pulse fluence and repetition rate of the laser, we increased scanning speed of the translation stage until the structure formation stops. From the obtained data we recalculate the number of pulses per spot which are necessary to create the structure for different pulse fluence and repetition rate (Fig. 1d). Since we know the thickness of Ti dioxide after LIPSS creation, we recalculate the height increase per pulse for different pulse fluence and repetition rate (red crosses on Fig. 1c).

\end{document}